\documentclass[preprint,showpacs,preprintnumbers,amsmath,amssymb]{revtex4}
\usepackage{amsmath,amssymb,graphics,epsfig,subfigure}

\begin{document}

\title{Fermion Localization and Resonances on A de Sitter Thick Brane}

\author{Yu-Xiao Liu\footnote{E-mail: iuyx@lzu.edu.cn},
        Jie Yang\footnote{Corresponding author. E-mail: yangjiev@lzu.edu.cn} ,
        Zhen-Hua Zhao\footnote{E-mail: zhaozhh02@gmail.com},
        Chun-E Fu\footnote{E-mail: fuche08@lzu.cn},
        Yi-Shi Duan\footnote{E-mail: ysduan@lzu.edu.cn}}
\affiliation{Institute of Theoretical Physics,
      Lanzhou University, Lanzhou 730000, People's Republic of China}

\begin{abstract}
In arXiv:0901.3543, the simplest Yukawa coupling
$\eta\bar{\Psi}\phi\chi\Psi$ was considered for a
two-scalar-generated Bloch brane model. Fermionic resonances for
both chiralities were obtained, and their appearance is related to
branes with internal structure. Inspired on this result, we
investigate the localization and resonance spectrum of fermions on
a one-scalar-generated $dS$ thick brane with a class of
scalar-fermion couplings $\eta\bar{\Psi}\phi^k\Psi$ with positive
odd integer $k$. A set of massive fermionic resonances for both
chiralities are obtained when provided large couple constant
$\eta$. We find that the masses and life-times of left and right
chiral resonances are almost the same, which demonstrates that it
is possible to compose massive Dirac fermions from the left and
right chiral resonances. The resonance with lower mass has longer
life-time. For a same set of parameters, the number of resonances
increases with $k$ and the life-time of the lower level resonance
for larger $k$ is much longer than the one for smaller $k$.
\end{abstract}

\pacs{ 11.10.Kk., 04.50.+h.}

\maketitle

\section{Introduction}

During the last two decades, the idea of embedding our universe in
a higher dimensional space has received much attention. The
suggestion that extra dimensions may not be compact
\cite{Akama1982,RubakovPLB1983136,VisserPLB1985,Randjbar-DaemiPLB1986,rs,Lykken}
or large \cite{AntoniadisPLB1990,ADD} can provide new insights for
solving several puzzling phenomena such as the gauge hierarchy
\cite{ADD}, i.e., the large difference in magnitude between the
Planck and electroweak scales, the dark matter origin and the
long-standing cosmological constant problem
\cite{RubakovPLB1983136,Randjbar-DaemiPLB1986,CosmConst,StojkovicCHM,ShtanovJCAP2009}.
According to the brane scenarios, gravity is free to propagate in
all dimensions, while all the matter fields (electromagnetic,
Yang-Mills etc.) are confined to a 3--brane in a larger
dimensional space. In Randall-Sundrum (RS) brane model \cite{rs},
an alternative scenario of the compactification had been proposed.
In this scenario, the internal manifold does not need to be
compactified to the Planck scale any more, it can be large, or
even infinite non-compact, which is one of the reasons why this
new compactification scenario has attracted so much attention.

Recently, an increasing interest has been focused on the study of
thick brane scenario in higher dimensional space-time
\cite{dewolfe,gremm,Csaki,CamposPRL2002,WangPRD2002,varios,ThickBrane,Sarrazin0903},
since in more realistic models the thickness of the brane should be
taken into account. In this scenario the scalar field provides a
thick brane realization of the brane world as a domain wall in the
bulk. For a comprehensive review on the thick brane solutions and
related topics please see Ref. \cite{ThickBraneReview}. However,
there are not so many analytic solutions of a dynamic thick domain
wall. The de Sitter ($dS$) branes have been studied in five and
higher dimensional spacetimes, for examples in
\cite{WangPRD2002,SasakuraJHEP2002,MinamNPB737,DzhunushalievPRD2009}.
The localization problem of spin half fermions on thick branes is
interesting and important. The problem of fermion localization and
the generation puzzle in 6D models were studied in Ref.
\cite{Singleton0607}. Usually, fermions do not have normalizable
zero modes in five dimensions without the scalar-fermion coupling
\cite{BajcPLB2000,Liu0708,Grossman2000,NonLocalizedFermion,IchinosePRD2002,Ringeval,RandjbarPLB2000,KoleyCQG2005,DubovskyPRD2000,0803.1458}.
In five dimensions, with the scalar--fermion coupling, there may
exist a single bound state and a continuous gapless spectrum of
massive fermion Kaluza-Klein (KK) states
\cite{Liu0708,ThickBraneWeyl}, while for some other brane models,
there exist finite discrete KK states (mass gap) and a continuous
gapless spectrum starting at a positive $m^2$
\cite{ThickBrane4,Liu0803}. In Ref. \cite{KoleyCQG2005}, the authors
obtained trapped discrete massive fermion states on the brane, which
are in fact quasi-bound and have a finite probability of escaping
into the bulk. In fact, fermions can escape into the bulk by
tunnelling, and the rate depends on the parameters of the scalar
potential \cite{DubovskyPRD2000}. In \cite{0901.3543}, the simplest
Yukawa coupling between two scalars and a spinor field was
considered for a two-scalar-generated Bloch brane model
\cite{BlochBranes}. Fermionic resonances for both chiralities were
obtained, and their appearance is related to branes with internal
structure.

In Ref. \cite{LiuJCAP2009}, we addressed the localization and mass
spectra of various bulk matter fields on the $dS$ thick branes. It
was shown that, all bulk matters (scalars, vectors and fermions) can
be localized on these branes and the corresponding mass spectra have
a mass gap. However, for spin half fermions the scalar-fermion
coupling should not be the usual Yukawa coupling $\eta\bar{\Psi}\phi
\Psi$ in order to trap the zero modes.
   In this paper, inspired on the results obtained in Refs.
\cite{DubovskyPRD2000,KoleyCQG2005,0901.3543}, we reinvestigate the
localization problem of fermions on a one-scalar-generated $dS$
thick brane
\cite{WangPRD2002,LiuJCAP2009,Goetz:1990,Gass:1999gk,Guerrero:2002ki}
with a class of scalar-fermion couplings $\eta\bar{\Psi}\phi^k\Psi$
with positive odd integer $k$. We shall show that, a set of massive
fermionic resonances for both chiralities could be obtained when
provided large couple constant $\eta$, and it is possible to compose
massive Dirac fermions from the left and right chiral resonances.

The organization of the paper is as follows: In section
\ref{SecModel}, we first review the one-scalar-generated $dS$ thick
brane in a 5-dimensional space-time. Then, in section
\ref{SecLocalize}, we study localization and mass spectra of
fermions on the thick brane by presenting the shapes of the
mass-independent potentials of the corresponding Schr\"{o}dinger
problem of fermionic KK modes. We consider two different types of
scalar-fermion interactions. Finally, the discussion and conclusion
are given in section \ref{conclusion}.

\section{One-scalar-generated de Sitter thick brane}
\label{SecModel}

Let us consider the de Sitter thick brane arising from a real scalar
field $\phi$ with a scalar potential $V(\phi)$. Our system is
described by the action
\begin{equation}
S = \int d^5 x \sqrt{-g}\left [ \frac{1}{2\kappa_5^2} R-\frac{1}{2}
g^{MN}\partial_M \phi
\partial_N \phi - V(\phi) \right ],
\label{action}
\end{equation}
where $\kappa_5^2=8 \pi G_5$ with $G_5$ the 5-dimensional Newton
constant. Here we set $\kappa_5=1$. The line-element for a
5-dimensional spacetime with planar-paralell symmetry is assumed as
\begin{eqnarray}
 ds^2&=&\text{e}^{2A}\big(\hat{g}_{\mu\nu}dx^\mu dx^\nu
          + dz^2\big)    \nonumber \\
 &=&\text{e}^{2A}\big(-dt^2+e^{2\beta t}dx^i dx^i + dz^2\big),
\label{linee}
\end{eqnarray}
where $z$ stands for the extra coordinate. The scalar field $\phi$
and the warp factor $\text{e}^{2A}$ are considered to be functions
of $z$ only, i.e., $\phi=\phi(z)$ and $A=A(z)$. In the model, the
potential could provide a thick brane realization. The field
equations generated from the action (\ref{action}) with the ansatz
(\ref{linee}) reduce to the following coupled nonlinear differential
equations
\begin{eqnarray}
\phi'^2 & = & 3(A'^2-A''-\beta^2), \label{coupledEqa} \\
V(\phi) & = & \frac{3}{2} e^{-2A}
 (-3A'^2-A''+3\beta^2), \label{coupledEqb}\\
\frac{dV(\phi)}{d\phi} &  = & e^{-2A}(3A'\phi'+\phi''),
\label{coupledEqc}
\end{eqnarray}
where the prime denotes derivative with respect to $z$. For positive
and vanishing ${\beta}$ one will obtain dynamic and static
solutions, respectively.

A de Sitter thick brane solution in a five-dimensional spacetime for
the potential
\begin{eqnarray}
V(\phi)=\frac{1+3\delta}{2\delta}\ 3\beta^{2}\left(\cos
\frac{\phi}{\phi_{0}} \right)^{2(1-\delta)}
          \label{potencial goetz}
\end{eqnarray}
was found in Refs. \cite{Goetz:1990,Gass:1999gk}:
\begin{eqnarray}
 e^{2A}&=&\cosh^{-2\delta}\left(\frac{\beta z}{\delta}\right) ,
         \label{e2A1} \\
 \phi~&=&\phi_{0}\arctan\left(\sinh \frac{\beta z}{\delta}\right),
         \label{phi1}
\end{eqnarray}
where $\phi_{0} =\sqrt{3\delta(1-\delta)}$, $0<\delta<1, \beta>0$.
The soliton configuration of the scalar field dynamically generates
the domain wall configuration with warped geometry. In this system,
The scalar field takes values $\pm\phi_0\pi/2$ at $z\rightarrow\pm
\infty$, corresponding to two consecutive minima of the potential
with cosmological constant $\Lambda=0$. The scalar configuration in
fact is a kink. The thick brane has a well-defined thin wall limit
when $\delta\rightarrow 0$ \cite{Guerrero:2002ki} and can localize
gravity and matter fields on the wall
\cite{WangPRD2002,LiuJCAP2009}. For $1/2 < \delta < 1$, the
hypersurfaces $|z|=\infty$ represent non-scalar spacetime
singularities \cite{WangPRD2002}. The energy density $\rho$ for the
$dS$ brane is calculated as follows:
\begin{eqnarray}
 \rho &=& 3e^{-2A}\big(\beta^2 -A'^2-A'' \big) \nonumber \\
  &=&\frac{3\beta^2(1+\delta)}{\delta}
  \cosh^{2(\delta-1)}\left(\frac{\beta z}{\delta}\right).
       \label{EnergyDensity}
\end{eqnarray}

\section{Fermion localization and resonances on the de Sitter thick brane}
\label{SecLocalize}

Fermions on branes have been studied in a number of articles such
as in Refs.
\cite{CasadioPRD2001,CasadioPLB2000,LiuPRD2008,LiuJHEP2007,Parameswaran0608074,LiuNPB2007,
TakamizuPLB2007,LiuVortexFermion,Koley,20082009,KodamaPRD2009}. In
this section let us investigate whether spin 1/2 fermions can be
localized on the thick brane given in previous section. The Dirac
action of a massless spin 1/2 fermion coupled to the background
scalar $\phi$ (\ref{phi1}) is
\begin{eqnarray}
S_{1/2} = \int d^5 x \sqrt{-g} \left(\bar{\Psi} \Gamma^M
          (\partial_M+\omega_M) \Psi
          -\eta \bar{\Psi} F(\phi)\Psi\right).~~ \label{DiracAction}
\end{eqnarray}
The non-vanishing components of the spin connection $\omega_M$ for
the background metric (\ref{linee}) are
\begin{eqnarray}
  \omega_\mu =\frac{1}{2}A' \gamma_\mu \gamma_5
             +\hat{\omega}_\mu, \label{spinConnection}
\end{eqnarray}
with $\hat{\omega}_\mu$ the spin connection derived from the metric
$\hat{g}_{\mu\nu}(x)$. Then the equation of motion is given by
\begin{eqnarray}
 \big[ \gamma^{\mu}(\partial_{\mu}+\hat{\omega}_\mu)
         + \gamma^5 \left(\partial_z  +2 A'\right)
         -\eta\; \text{e}^A F(\phi)
 \big ] \Psi =0. \label{DiracEq1}
\end{eqnarray}
Note that the sign of the coupling $\eta$ of the spinor $\Psi$ to
the scalar $\phi$ is arbitrary and represents a coupling either to
kink or to anti-kink domain wall. For definiteness, we shall
consider in what follows only the case of a kink coupling, and thus
assume that $\eta>0$.

By making the following general chiral decomposition:
\begin{equation}
 \Psi = \sum_n\big[\psi_{L,n}(x) f_{L,n}(z)
 +\psi_{R,n}(x) f_{R,n}(z)\big]\text{e}^{-2A}
\end{equation}
with $\psi_{L}=-\gamma^5 \psi_{L}$ and $\psi_{R}=\gamma^5 \psi_{R}$
the left-handed and right-handed components of a 4D Dirac field
respectively, and demanding $\psi_{L,R}$ satisfy the 4D massive
Dirac equations
$\gamma^{\mu}(\partial_{\mu}+\hat{\omega}_\mu)\psi_{L,R}
=m\psi_{R,L}$, we obtain the following coupled equations
\begin{subequations}\label{CoupleEq1}
\begin{eqnarray}
 \left[\partial_z + \eta\;\text{e}^A F(\phi) \right]f_L(z)
  &=&  ~~m f_R(z), \label{CoupleEq1a}  \\
 \left[\partial_z- \eta\;\text{e}^A F(\phi) \right]f_R(z)
  &=&  - m f_L(z), \label{CoupleEq1b}
\end{eqnarray}
\end{subequations}
which can be reduced to the Schr\"{o}dinger-like equations for the
wavefunctions of left and right chiral fermions
\begin{subequations}\label{SchEqFermion}
\begin{eqnarray}
  \big(-\partial^2_z + V_L(z) \big)f_{L}
            &=&m^2 f_{L},~~
   \label{SchEqLeftFermion}  \\
  \big(-\partial^2_z + V_R(z) \big)f_{R}
            &=&m^2 f_{R},
   \label{SchEqRightFermion}
\end{eqnarray}
\end{subequations}
where the effective potentials are given by
\begin{subequations}\label{Vfermion}
\begin{eqnarray}
  V_L(z)&=& \big(\eta \text{e}^{A} F(\phi)\big)^2
     -\partial_z \big(\eta \text{e}^{A} F(\phi)\big), \label{VL}\\
  V_R(z)&=&   V_L(z)|_{\eta \rightarrow -\eta}. \label{VR}
\end{eqnarray}
\end{subequations}
We have dropped the index $n$ for convenience.

It can be seen that, in order to localize left or right chiral
fermions, there must be some kind of scalar-fermion coupling, and
the effective potential $V_L(z)$ or $V_R(z)$ should have a minimum
at the location of the brane. Furthermore, for the kink
configuration of the scalar $\phi(z)$ (\ref{phi1}), $F(\phi(z))$
should be an odd function of $\phi(z)$ when one demands that
$V_{L,R}(z)$ are invariant under $Z_2$ reflection symmetry
$z\rightarrow -z$. Thus we have $F(\phi(0))=0$ and
$V_L(0)=-V_R(0)=-\eta\partial_z F(\phi(0))$, which results in the
well-known conclusion: only one of the massless left and right
chiral fermions could be localized on the brane. The spectra are
determined by the behavior of the potentials at infinity. For
$V_{L,R}\rightarrow 0$ as $|z|\rightarrow \infty$, one of the
potentials would have a volcano-like shape and there exists only a
bound massless mode followed by a continuous gapless spectrum of KK
states, while another could not trap any bound states and the
spectrum is also continuous and gapless. For $V_{L,R}\rightarrow
V_{\infty}=$ const as $|z|\rightarrow \infty$, those modes with
$m_n^2<V_{\infty}$ belong to discrete spectrum and modes with
$m_n^2>V_{\infty}$ contribute to continuous one. If the potentials
increase as $|z|\rightarrow \infty$ the spectra are discrete.

The concrete behavior of the potentials is depended on the function
$F(\phi)$. In Ref. \cite{LiuJCAP2009}, two cases $F(\phi)=\phi$ and
$F(\phi)=\sin(
\frac{\phi}{\phi_0})\cos^{-\delta}(\frac{\phi}{\phi_0})$ were
investigated in detail as examples. For the first example with usual
Yukawa couple $\eta\bar{\Psi}\phi\Psi$, there exists no mass gap but
a continuous gapless spectrum of KK states. For the second example
with a positive coupling constant $\eta$, there exist some discrete
bound KK modes and a series of continuous ones. The total number of
bound states increases with the couple constant $\eta$. If
$0<\eta<\beta/\delta$, there is only one left chiral fermion bound
state which is just the left chiral fermion zero mode; if
$\eta>\beta/\delta$, there are $N_{max}+1$ left chiral fermion bound
states (including zero mode and massive KK modes) and $N_{max}$
right chiral fermion bound states (including only massive KK modes).
Especially, the spectrum of left chiral fermions
\begin{eqnarray}
 &&m^2_{L,n} = \frac{\beta}{\delta^2}{(2\delta\eta-n\beta)n}~~~\nonumber \\
 &&(\eta>0,~n=0,1,2,...,N_{max}<\frac{\delta\eta}{\beta})~~
 \label{massSpectrumVL}
\end{eqnarray}
and right one
\begin{eqnarray}
 &&m^2_{R,n} = \frac{\beta}{\delta^2}
   {\big[2\delta\eta- (n+1)\beta\big](n+1)}~~~\nonumber \\
   &&(\eta> \frac{\beta}{\delta},~
    n=0,1,2,...,N_{max}-1<\frac{\delta\eta}{\beta}-1)~~
 \label{massSpectrumVR}
\end{eqnarray}
are the same except the massless mode, i.e.,
$m^2_{L,n+1}=m^2_{R,n}$. In this paper, we would like to consider
the case $F(\phi)=\phi^k$ with $k$ a positive odd integer, for the
purpose of investigating resonances of massive fermions.

\subsection{Case I: $F(\phi)=\phi$}

\begin{figure}[htb]
\begin{center}
\includegraphics[width=7cm,height=4.5cm]{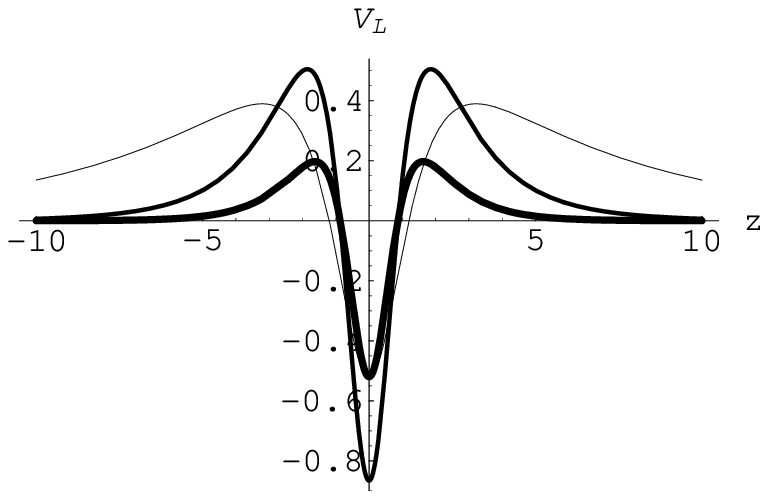}
\includegraphics[width=7cm,height=4.5cm]{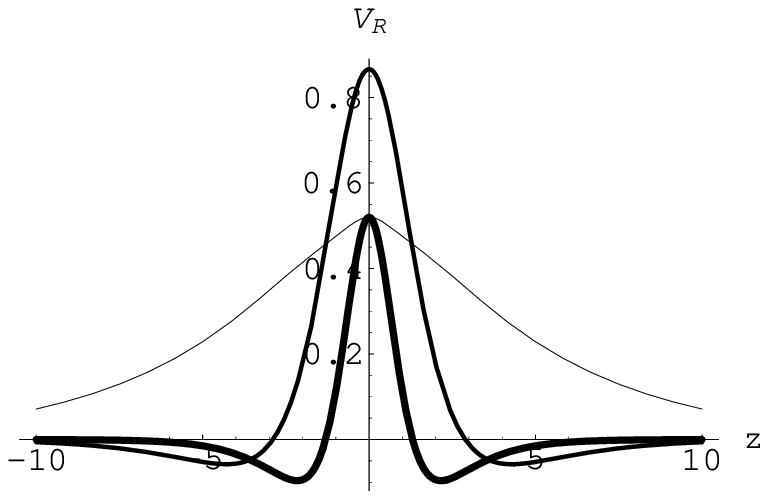}
\end{center} \vskip -3mm
\caption{The potentials $V_{L,R}$ for left and right chiral fermions
with $F(\phi)=\phi$. The parameters are set to $\eta=1$ and
$\delta=\beta=0.1, 0.5, 0.9$ for lines with thickness increases with
$\delta$.}
 \label{fig:VLR}
\end{figure}

\begin{figure}[htb]
\begin{center}
 \subfigure[$\eta=3$]  {\label{fig:VReta3}
\includegraphics[width=7cm,height=4.5cm]{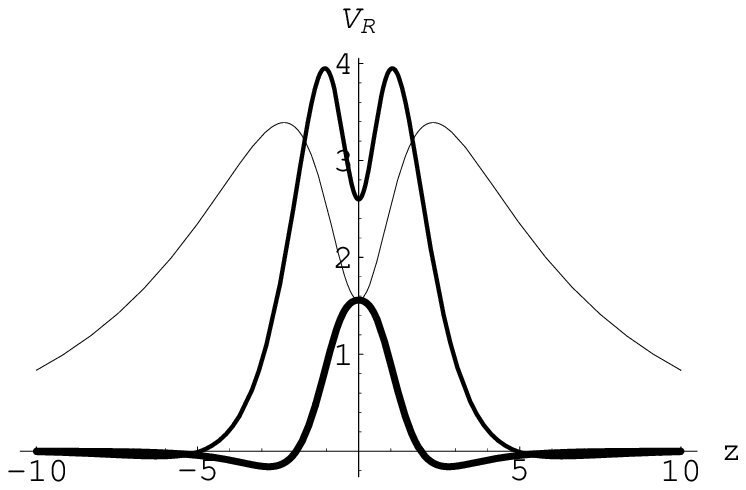}}
 \subfigure[$\eta=10$]  {\label{fig:VReta10}
\includegraphics[width=7cm,height=4.5cm]{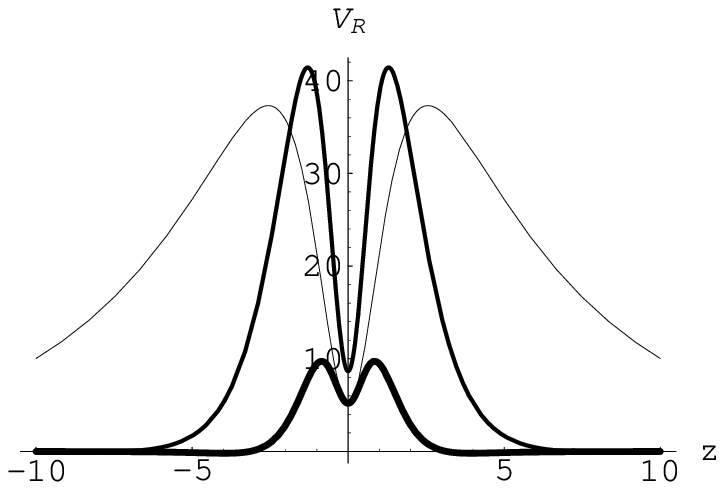}}
\end{center}\vskip -5mm
\caption{The potential $V_{R}$ for right chiral fermions with
$F(\phi)=\phi$. The parameters are set to $\eta=3$ and $\eta=10$,
$\delta=\beta=0.1, 0.5, 0.9$ for lines with thickness increases with
$\delta$.}
 \label{fig:VR}
\end{figure}

Firstly, we reconsider the simplest case $F(\phi)=\phi$ for the $dS$
brane world solution (\ref{potencial goetz})-(\ref{phi1}). The
explicit forms of the potentials (\ref{Vfermion}) are
\begin{subequations}\label{VSLRphiCaseI}
\begin{eqnarray}
 V_L(z)
  &=& \eta^2 \phi_0^2
     \frac{ \arctan^2 \sinh \left( \frac{\beta z}{\delta} \right)}
          {\cosh^{2\delta}\left( \frac{\beta z}{\delta}  \right) }
     -\frac{\eta\beta\phi_0}{\delta\cosh^{1+\delta}\left(\frac{\beta z}{\delta}\right)}
      \nonumber \\
  &&  +\eta\beta\phi_0
    \frac{\sinh \left( \frac{\beta z}{\delta} \right)
            \arctan \sinh \left( \frac{\beta z}{\delta} \right)}
         {\cosh^{1+\delta}\left(\frac{\beta z}{\delta}\right)}
      ,  \label{VSLphiCaseI} \\
  V_R(z) &=& V_L(z)|_{\eta \rightarrow -\eta}. \label{VSRphiCaseI}
\end{eqnarray}
\end{subequations}
The values of the potentials (\ref{VSLphiCaseI}) and
(\ref{VSRphiCaseI}) at $z = 0$ and $z \rightarrow \pm\infty$ are
given by
\begin{eqnarray}
&&V_L(0) =-V_R(0) = -\beta\eta
         \sqrt{3(\delta^{-1}-1)}\;, \\
&&V_{L,R}(z\rightarrow\pm \infty)\rightarrow0,
\end{eqnarray}
i.e., both potentials have same asymptotic behavior when $z
\rightarrow \pm\infty$, but opposite behavior at the origin $z=0$.
In this paper we take $\beta=\delta$ and $\eta>0$ for simplicity.
The shapes of the potentials are shown in Figs. \ref{fig:VLR} and
\ref{fig:VR}. It can be seen that, for any $0<\delta<1$ and
$\eta>0$, $V_L(z)$ is indeed a modified volcano type potential.
Hence, the potential of left chiral fermions provides no mass gap to
separate the zero mode from the excited KK ones, and there exists a
continuous gapless spectrum of the KK modes for left chiral
fermions. However, the zero mode of the left chiral fermions
\begin{equation}
 f_{L0}(z)  \propto \exp\left(-\eta\int^z d\bar{z}
   \text{e}^{A(\bar{z})}\phi(\bar{z})\right)
  \label{zeroModeL}
\end{equation}
is non-normalizable \cite{LiuJCAP2009}, which is an example that
negative value of potential at the brane location does not
guarantee the existence of a normalized zero mode. This is
different from the situation in Refs.
\cite{KoleyCQG2005,LiuPRD2008}, where the corresponding potential
of left chiral fermions is also a volcano type one and the zero
mode can be localized on the branes provided strong enough Yukawa
coupling.

The potential $V_R$ is always positive at the brane location and
vanishes when far away from the brane. This shows that the
potential could not trap any bound fermions with right chirality
and there is no zero mode of right chiral fermions. This agrees
with the well-known fact that massless fermions must be
single-handed in brane models \cite{Ringeval}. However, the shape
of the potential is depended on the coupling constant $\eta$. With
the increase of $\eta$, a potential well around the brane location
would appear and the well becomes more and more deeper. The
appearance of the potential well could be related to resonances or
to massive fermions with a finite life-time. In Ref.
\cite{0901.3543}, a similar potential and resonances for left and
right chiral fermions were found but in background of two-field
thick branes with internal structure. In what follows, we follow
Ref. \cite{0901.3543} and investigate the massive modes of
fermions by solving numerically Eq. (\ref{SchEqFermion}) with
potentials in (\ref{VSLRphiCaseI}). Differently, we present
another method to calculate the probability for finding the
massive modes on the brane.

\begin{figure}[htb]
\begin{center}
\subfigure[$m^2=15.9$] {\label{fig:fLm2.1}
\includegraphics[width=7cm,height=4.5cm]{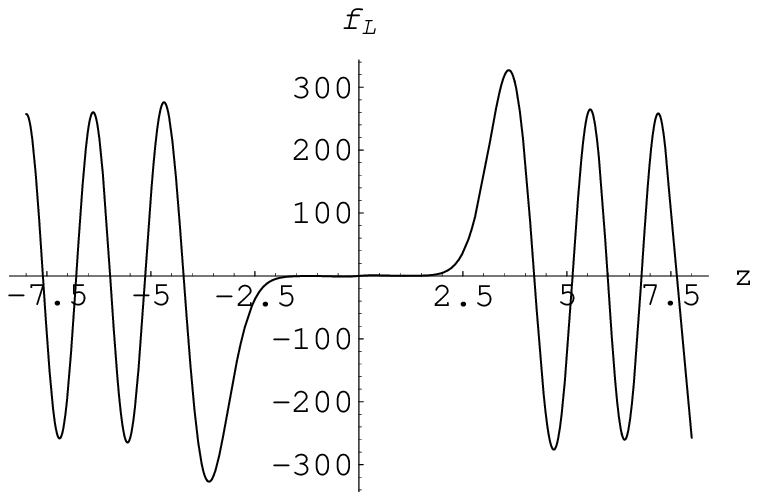}}
\subfigure[$m^2=16$] {\label{fig:fLm2.2}
\includegraphics[width=7cm,height=4.5cm]{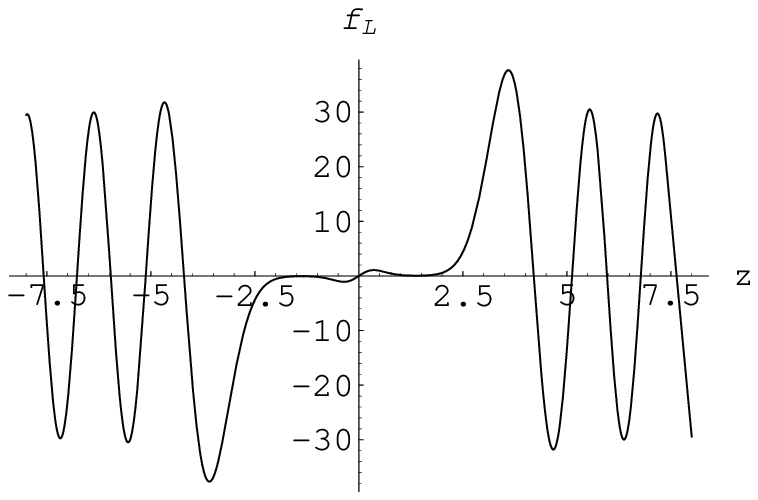}}
\subfigure[$m^2=16.013$] {\label{fig:fLm2.3}
\includegraphics[width=7cm,height=4.5cm]{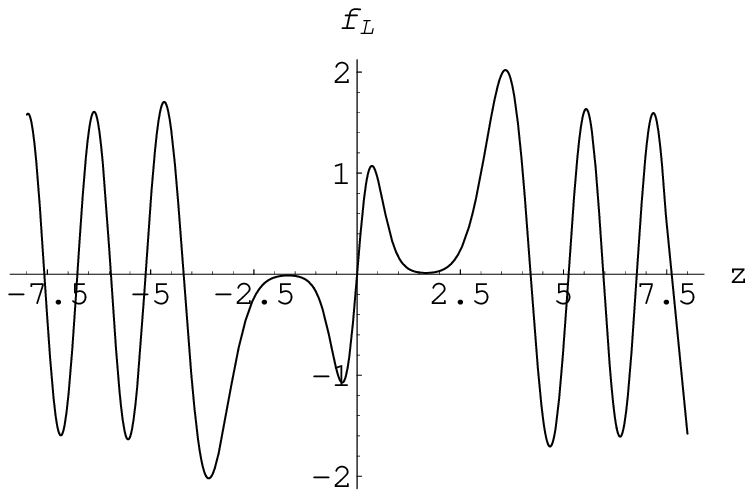}}
\subfigure[$m^2=16.0137$] {\label{fig:fLm2.4}
\includegraphics[width=7cm,height=4.5cm]{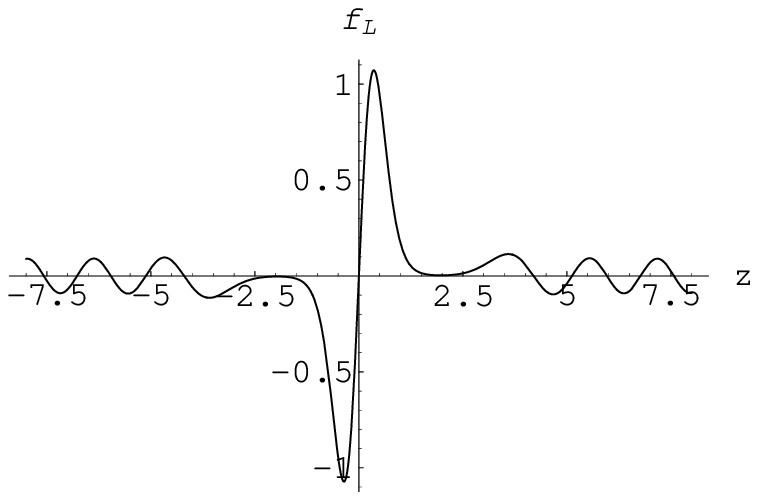}}
\subfigure[$m^2=16.01372$] {\label{fig:fLm2.5}
\includegraphics[width=7cm,height=4.5cm]{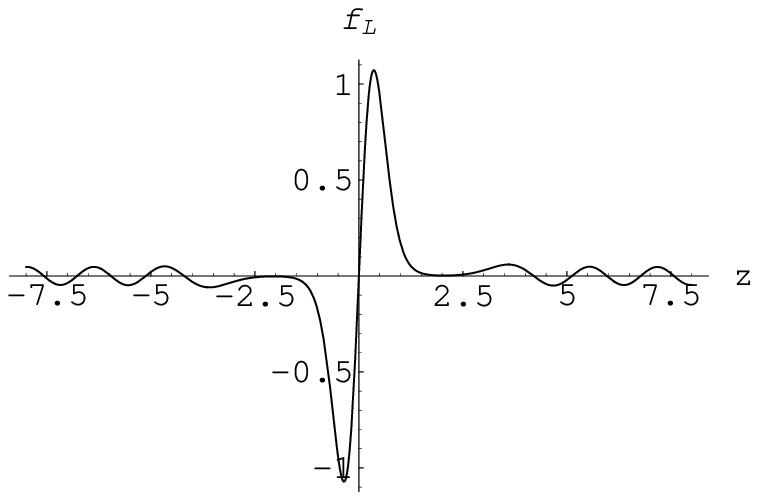}}
\subfigure[$m^2=16.01374$] {\label{fig:fLm2.6}
\includegraphics[width=7cm,height=4.5cm]{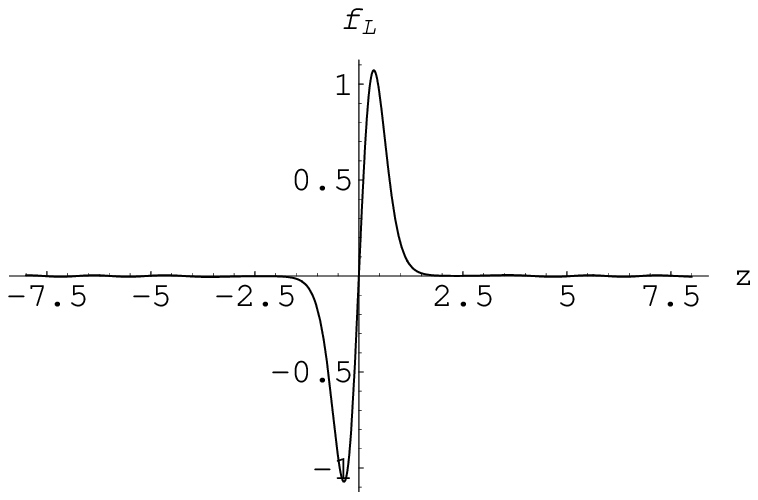}}
\end{center}\vskip -6mm
\caption{Massive KK modes of left chiral fermions for the case
$F(\phi)=\phi$ with different $m^2$. The parameters are
$\delta=\beta=0.5$ and $\eta=10$.}
 \label{fig:fLm2}
\end{figure}

In order to get the solutions of KK modes $f_{L,R}(z)$ from the
second order differential equations (\ref{SchEqFermion}), we need to
impose two initial conditions. The wavefunctions of a
Schr\"{o}dinger equation with a finite smooth potential are
continuous at any position. Furthermore, considering the even parity
of the potentials, we can impose two kinds of initial conditions:
\begin{equation}
 f(0)=c_0, f'(0)=0,   \label{initialCondition1}
\end{equation}
and
\begin{equation}
 f(0)=0, f'(0)=c_1.   \label{initialCondition2}
\end{equation}
The first and second conditions would result in even and odd KK
modes, respectively. The constants $c_0$ and $c_1$ for unbound
massive KK modes are arbitrary but will be set to $c_0=1$ and
$c_1=5$. The massive KK modes would encounter the tunneling process
across the potential barriers near the brane. And the modes with
different masses would have different life-times. Some massive KK
modes of left chiral fermions for the case $F(\phi)=\phi$ with
different $m^2$ are plotted in Fig. \ref{fig:fLm2}. These shapes
show that there could exist some resonant states at some $m^2$. In
what follows, we would like to investigate this problem.

\begin{figure}[htb]
\includegraphics[width=7cm,height=4.5cm]{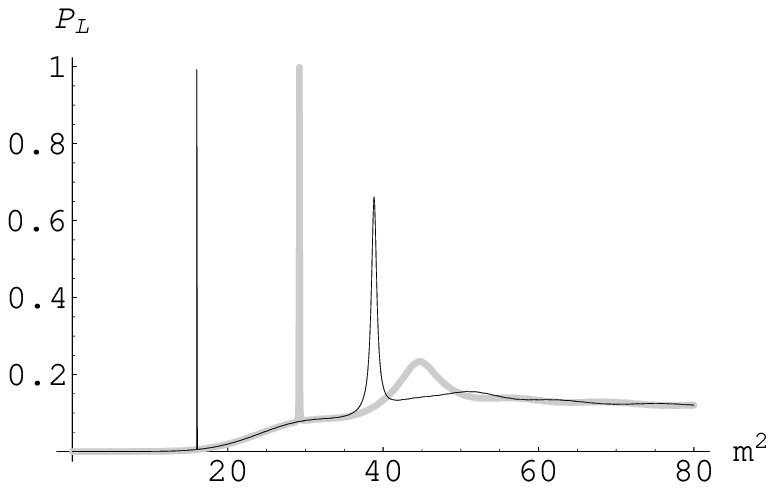}
\includegraphics[width=7cm,height=4.5cm]{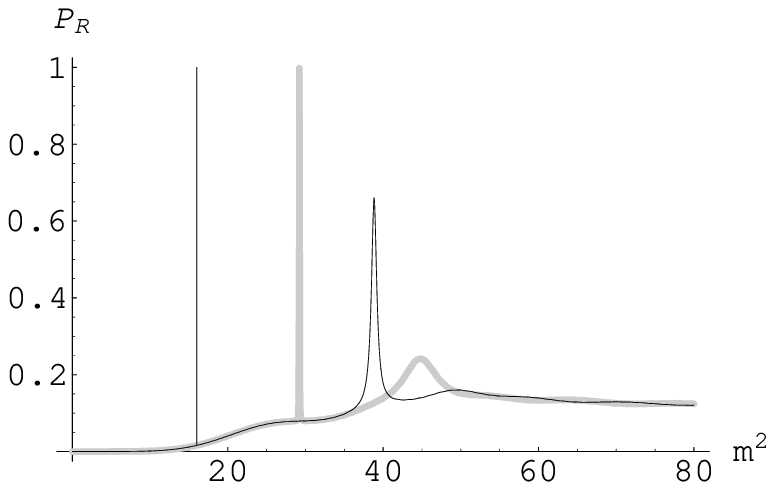}
\includegraphics[width=7cm,height=4.5cm]{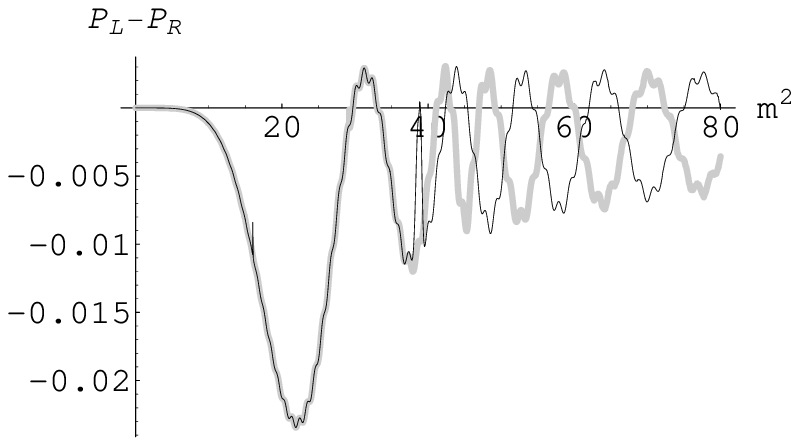}
\caption{The probability for finding massive KK modes of left and
right chiral fermions with mass $m^2$ around the brane location,
$P_{L,R}$, and their differences $P_L-P_R$, as a function of $m^2$,
for the case $F(\phi)=\phi$. The parameters are $\delta=\beta=0.5$
and $\eta=10$. For $P_L$, gray thick lines and black thin lines are
plotted for the first and second initial condition, respectively.
For $P_R$, black thin lines and gray thick lines are plotted for the
first and second initial condition, respectively.}
 \label{fig:PLR}
\end{figure}

Since Eq. (\ref{SchEqFermion}) can be rewritten as
${\cal{O}}_{L,R}^\dag {\cal{O}}f_{L,R}(z)=m^2 f_{L,R}(z)$, one can
interpret $|f_{L,R}(z)|^2$ as the probability for finding the
massive KK modes at the position $z$ along extra dimension
\cite{0901.3543}. In Ref. \cite{0901.3543}, the authors suggested
that large peaks in the distribution of $f_{L,R}(0)$ as a function
of $m$ would reveal the existence of resonant states. Here we need
to extend this idea for our two kinds of initial conditions
(\ref{initialCondition1}) and (\ref{initialCondition2}). This is
because the value of $f_{L,R}(0)$ is zero for condition
(\ref{initialCondition2}). We proposal that large relative
probabilities for finding massive KK modes within a narrow range
$-z_b<z<z_b$ around the brane location, are called $P_{L,R}$, would
indicate the existence of resonances. We can consider the KK modes
$f_{L,R}(z)$ in a box with borders $|z|=z_{max}$ located far from
the turning point, beyond which $f_{L,R}(z)$ are turned into plane
waves. The relative probabilities are defined as follows:
\begin{equation}
 P_{L,R}(m)=\frac{\int_{-z_b}^{z_b} |f_{L,R}(z)|^2 dz}
                 {\int_{-z_{max}}^{z_{max}} |f_{L,R}(z)|^2 dz},
 \label{Probability}
\end{equation}
where we choose $z_b=0.1 z_{max}$, for which the probability for a
plane wave mode with mass $m$ would be 0.1. For these KK modes with
more larger $m^2$ than the maximum of the corresponding potential,
they would be approximatively plane waves and the probabilities for
them would trend to 0.1. For $\delta=\beta=0.5$ and $\eta=10$, we
set $z_{max}=30$ and the results are shown in Fig. \ref{fig:PLR}. We
find from the figure a series of huge peaks located at
$m^2=16.013742$, $29.22241$ and $38.837314$ for both left-handed and
right-handed fermions. These peaks are related with resonances of
fermions, which are long-lived massive fermionic modes on the brane.
Except several peaks, the curves seemed to grow at first, and then
it plateaued around $z_b/z_{max}=0.1$. The reason is that, KK modes
with small $m^2 (\ll V_{L,R}^{max})$ will be damped near the brane
and oscillate away from the brane, while those modes with large $m^2
(\gg V_{L,R}^{max})$ can be approximated as plane wave modes
$f_{L,R}\propto\cos mz$ or $\sin mz$.

\begin{figure}[htb]
\begin{center}
\subfigure[$n=1,m^2=16.013742$] {\label{fig:ResonancesLeft1}
\includegraphics[width=7cm,height=4.5cm]{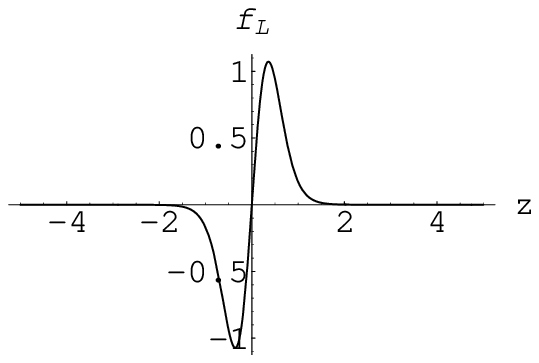}}
\subfigure[Zoom in on (a)] {\label{fig:ResonancesLeft2}
\includegraphics[width=7cm,height=4.5cm]{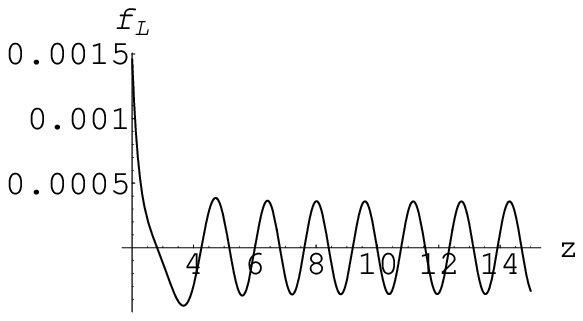}}
\subfigure[$n=2,m^2=29.22241$] {\label{fig:ResonancesLeft3}
\includegraphics[width=7cm,height=4.5cm]{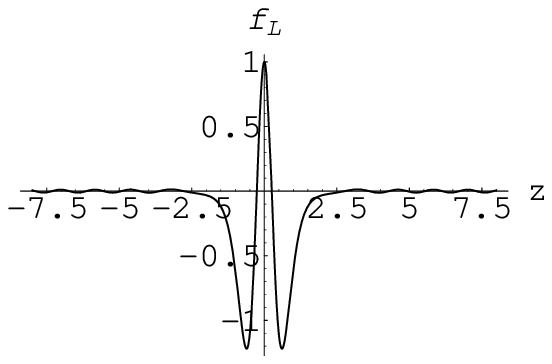}}
\subfigure[$n=3,m^2=38.837314$] {\label{fig:ResonancesLeft4}
\includegraphics[width=7cm,height=4.5cm]{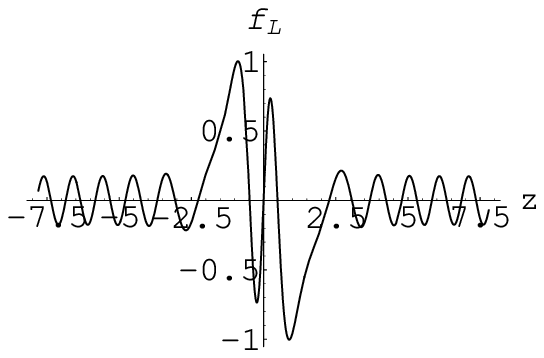}}
\end{center}\vskip -4mm
\caption{Massive KK modes of left chiral fermions $f_L(z)$ for the
case $F(\phi)=\phi$. The parameters are $\delta=\beta=0.5$ and
$\eta=10$.}
 \label{fig:ResonancesLeft}
\end{figure}

\begin{figure}[htb]
\begin{center}
\subfigure[$n=0,m^2=16.013742$] {\label{fig:ResonancesRight1}
\includegraphics[width=7cm,height=4.5cm]{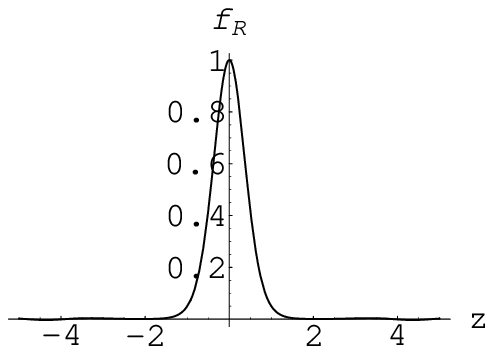}}
\subfigure[Zoom in on (a)] {\label{fig:ResonancesRight2}
\includegraphics[width=7cm,height=4.5cm]{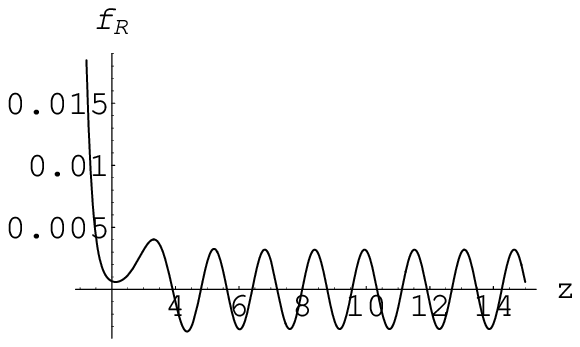}}
\subfigure[$n=1,m^2=29.22241$] {\label{fig:ResonancesRight3}
\includegraphics[width=7cm,height=4.5cm]{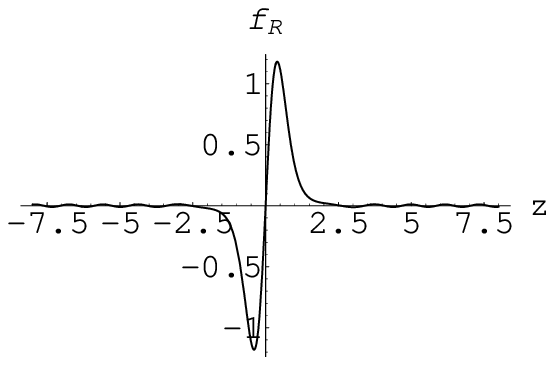}}
\subfigure[$n=2,m^2=38.837314$] {\label{fig:ResonancesRight4}
\includegraphics[width=7cm,height=4.5cm]{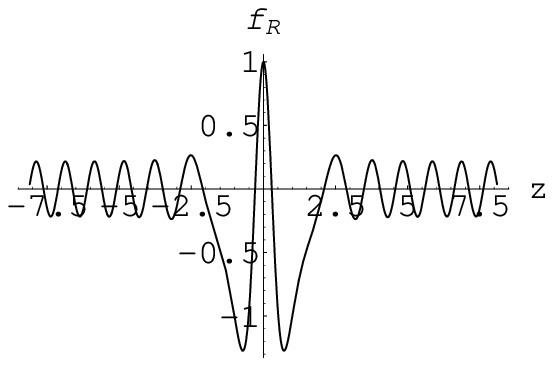}}
\end{center}\vskip -4mm
\caption{Massive KK modes of right chiral fermions $f_R(z)$ for the
case $F(\phi)=\phi$. The parameters are $\delta=\beta=0.5$ and
$\eta=10$.}
 \label{fig:ResonancesRight}
\end{figure}

In Figs. \ref{fig:ResonancesLeft} and \ref{fig:ResonancesRight}, we
plot the shapes of the resonances of left and right chiral fermions
for $\delta=\beta=0.5$ and $\eta=10$. It can be seen that the
configurations of Figs. \ref{fig:ResonancesLeft1},
\ref{fig:ResonancesLeft3} and \ref{fig:ResonancesLeft4} could
present the $n=1$, $n=2$ and $n=3$ level KK modes of left chiral
fermions, which are in fact resonances. The $n=0$ level mode
(\ref{zeroModeL}) with left chirality, the only one bound state, is
not shown here. While the configurations of Figs.
\ref{fig:ResonancesRight1}, \ref{fig:ResonancesRight3} and
\ref{fig:ResonancesRight4} present the $n=0$, $n=1$ and $n=2$ level
resonances of right chiral fermions. It is worth noting that the $n$
level massive resonance with left chirality and the $n-1$ level one
with right chirality have the same mass, i.e., the spectra of
massive left-handed and right-handed fermionic resonances are the
same. This demonstrates that it is possible to compose a Dirac
fermion from the left and right KK modes. If we only consider one of
the initial conditions (\ref{initialCondition1}) and
(\ref{initialCondition2}), then we could not get the resonances of
left-handed and right-handed fermions with same mass at the same
time. In fact, as mentioned before, we found in Refs.
\cite{LiuJCAP2009,Liu0803} that the spectra of bound massive KK
modes of left and right chiral fermions are the same, where the
effective potentials for KK modes of fermions are modified
P\"{o}schl-Teller potentials.

\begin{figure}[htb]
\begin{center}
\includegraphics[width=7cm,height=4.5cm]{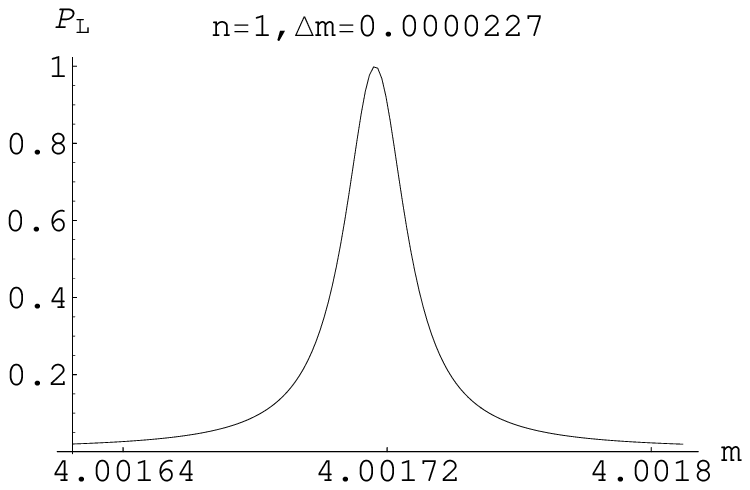}
\includegraphics[width=7cm,height=4.5cm]{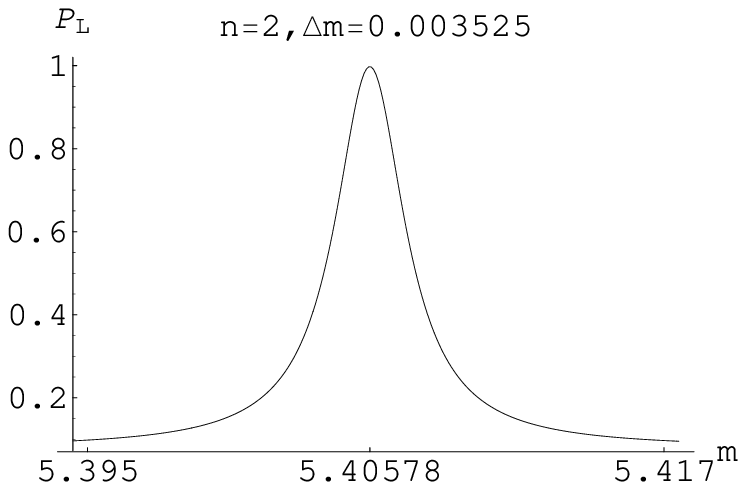}
\includegraphics[width=7cm,height=4.5cm]{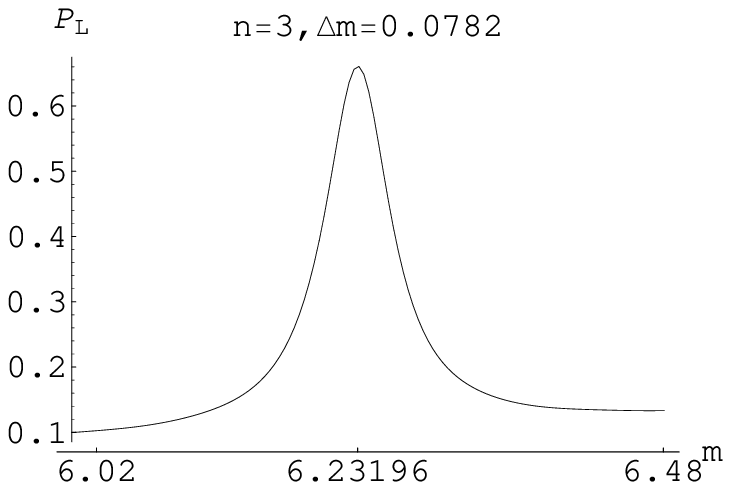}
\end{center}\vskip -4mm
\caption{The probability for finding massive KK modes of left
chiral fermions around the brane location, $P_{L}$, as a function
of $m$. The parameters are $\delta=\beta=0.5$ and $\eta=10$.}
 \label{fig:PLm}
\end{figure}

We can estimate the life-time $\tau$ for a resonance from the width
in mass $\Gamma=\delta m$ at half maximum of the corresponding peak
in Fig. \ref{fig:PLm}. This means that the fermion disappears into
the extra dimension with time $\tau\sim\Gamma^{-1}$
\cite{RubakovPRL2000}. The peaks corresponding to the resonances
shown in Figs. \ref{fig:ResonancesLeft} and
\ref{fig:ResonancesRight} are located at $m=4.00172$, $5.40578$ and
$6.23196$, respectively. And the width and life-time of the
resonances are listed in Table \ref{tab1}. It can be seen that the
resonances with lower mass have longer life-time. And the life-times
of left and right chiral resonances are almost the same, which
confirms further that it is possible to compose a Dirac fermion from
the left and right KK modes.

\begin{table}[h]
\begin{center}
\caption{The mass, width and life-time for resonances of left and
right chiral fermions. The parameters are $\delta=\beta=0.5$ and
$\eta=10$.}\label{tab1}
\renewcommand\arraystretch{1.3}
\begin{tabular}
 {|l||c|c|c|c|}
  \hline
  $~$ & $m^2$ & $m$ & $\Gamma$ & $\tau$  \\
  \hline\hline
  $n=1$(left) & 16.013742 & 4.00172 & 0.0000227 &  44052.9  \\
  \hline
  $n=0$(right) & 16.013742 & 4.00172 & 0.0000228 & 43859.6 \\
  \hline
  $n=2$(left) & 29.22241  & 5.40578 & 0.003525 &  283.688   \\
  \hline
  $n=1$(right) & 29.22241 & 5.40578 & 0.003534 & 282.965    \\
  \hline
  $n=3$(left) & 38.837314 & 6.23196  & 0.0782 & 12.7877  \\
  \hline
  $n=2$(right) & 38.837314 & 6.23196 & 0.0804 & 12.4378\\
  \hline
\end{tabular}
\end{center}
\end{table}

In order to get the resonances of massive KK modes for general
parameters $\delta,\beta$ and $\eta$, we expand the effective
potentials (\ref{VSLRphiCaseI}) around $z = 0$ and retain terms up
to order $z^2$. The differential equation for left chiral fermions
is reduced to
\begin{eqnarray}
  \big[\partial^2_z - (-c_0 - m^2 + c_2 z^2)
  \big]f_{L}(z)=0, \label{VL2order}
\end{eqnarray}
where
\begin{eqnarray}
 && c_0 = \frac{\beta}{\delta}\sqrt{3(1-\delta)\delta}~\eta, \label{c0}\\
 && c_2 = \frac{1}{2}\frac{\beta^3}{\delta^3}(1+3\delta)
            \sqrt{3(1-\delta)\delta}~\eta
          +\frac{\beta^2}{\delta^2}3(1-\delta)\delta\eta^2.~~~~~~~ \label{B2}
\end{eqnarray}
This is the harmonic oscillator approximation in the neighborhood of
the brane and the solution is
\begin{eqnarray}
 f_{L,n}(z) \propto \mathrm{e}^{-\frac{1}{2} \sqrt{c_2}\; z^2}
       H_n \big(c_2^{1/4} z\big),\label{SolutionOfScalar}
\end{eqnarray}
here $n=\frac{m^2+c_0}{2\sqrt{c_2}}-\frac{1}{2}$ is a nonnegative
integer and $H_n(z)$ are the Hermite polynomials. The possible
values of $m^2$ are given by
\begin{eqnarray}
  m_{L,n}^2&=&\frac{1+2n}{\sqrt{2}}
          \sqrt{\frac{\beta^2}{\delta^3}
                \left(6(1-\delta)\delta^2\eta
                  + \beta \sqrt{3(1-\delta)\delta}\;(1+3\delta)
                \right)\eta}  \nonumber \\
       &&  -\frac{\beta}{\delta} \sqrt{3(1-\delta)\delta}\; \eta
          ~~~~~(n=0,1,2,\cdots).
\end{eqnarray}
For right chiral fermions, we have
\begin{eqnarray}
  m_{R,n}^2&=&\frac{1+2n}{\sqrt{2}}
          \sqrt{\frac{\beta^2}{\delta^3}
                \left(6(1-\delta)\delta^2\eta
                    -\beta \sqrt{3(1-\delta)\delta}\;(1+3\delta)
                \right)\eta}  \nonumber \\
       &&  +\frac{\beta}{\delta} \sqrt{3(1-\delta)\delta}\; \eta
          ~~~~~(n=0,1,2,\cdots).
\end{eqnarray}
Note that, this is only an approximate solution for the spectra of
left and right chiral fermions. For zero mode of left chiral
fermions, we should have $m_{L,0}^2=0$, which results in the
constrained condition for the approximate solution:
\begin{eqnarray}
 \eta \gg \frac{\beta \sqrt{(1-\delta)\delta}\;(1+3\delta)}
          {2\sqrt{3}(1-\delta)\delta^2}. \label{ConstrainCondition}
\end{eqnarray}
If we take $\beta=\delta=0.5$, then we need to consider $\eta\gg
1.44$. Under the above condition, the mass spectra of fermions are
reduced to the following form:
\begin{eqnarray}
  m_{L,n}^2&=&2\frac{\beta}{\delta}
        \sqrt{3(1-\delta)\delta}\; \eta\; n,\\
  m_{R,n}^2&=&2\frac{\beta}{\delta}
        \sqrt{3(1-\delta)\delta}\; \eta\; (1+n).
\end{eqnarray}
Now, it is clear that the spectra of left and right massive fermions
are the same and only lower $n$ are available. The mass of the first
resonance of fermions is read as $m_{0}^2 =
2\sqrt{3(1-\delta)\delta}\; \eta{\beta}/{\delta}$.

It should point out that in order to get a better approximate
solution for the resonance spectra and KK modes of left and right
chiral fermions, we need to solve the Schr\"{o}dinger equation for
the approximate potential at both large $z$ and small $z$, and match
two solution to get the spectrum. In this paper, we do not discuss
the question in detail.

\subsection{Case II: $F(\phi)=\phi^k$ with odd $k>1$}

\begin{figure}[htb]
\begin{center}
\includegraphics[width=7cm,height=4.5cm]{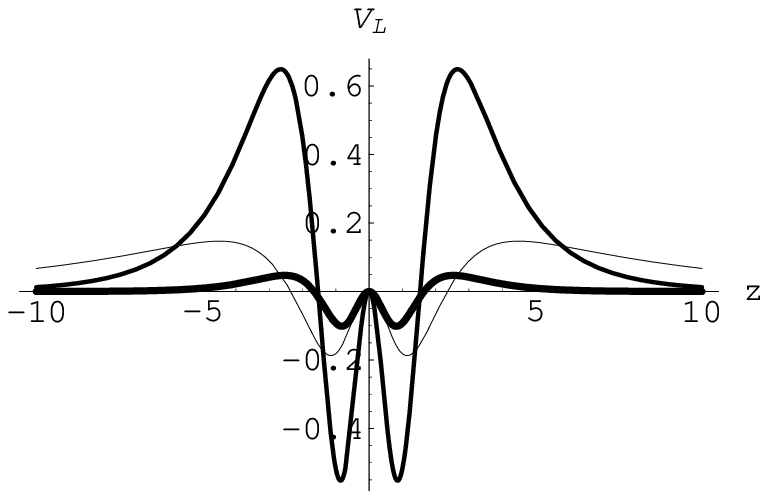}
\includegraphics[width=7cm,height=4.5cm]{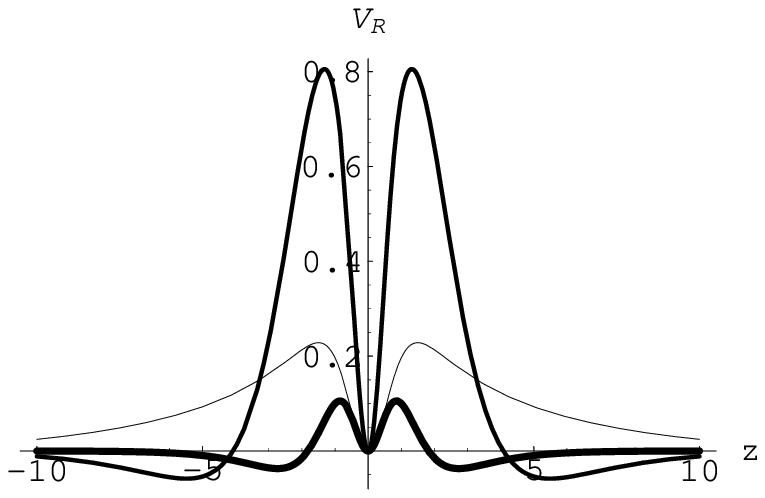}
\end{center} \vskip -4mm
\caption{The potentials $V_{L,R}$ for left and right chiral
fermions with $F(\phi)=\phi^3$. The parameters are set to $\eta=1$
and $\delta=\beta=0.1, 0.5, 0.9$ for lines with thickness
increases with $\delta$.}
 \label{fig:VLRCaseII}
\end{figure}

\begin{figure}[htb]
\begin{center}
\includegraphics[width=7cm,height=4.5cm]{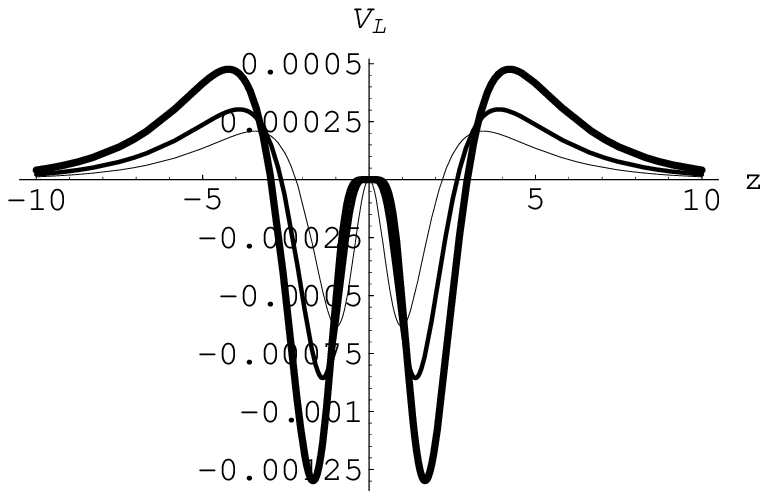}
\includegraphics[width=7cm,height=4.5cm]{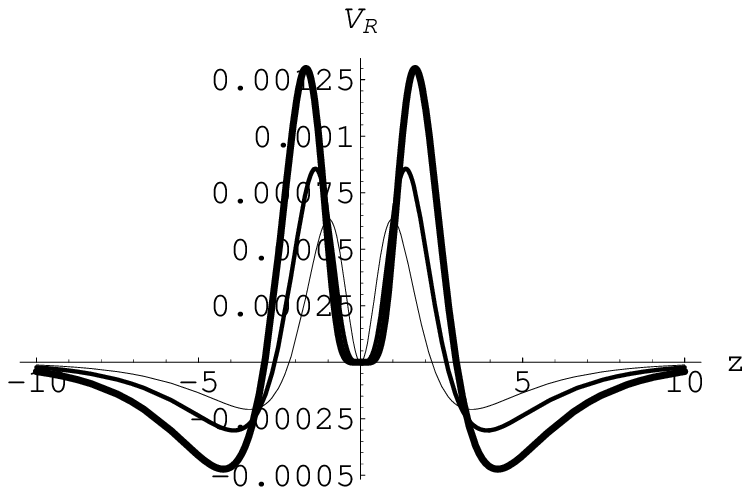}
\end{center} \vskip -4mm
\caption{The potentials $V_{L,R}$ for left and right chiral
fermions with $F(\phi)=\phi^k$. The parameters are set to
$\eta=0.001$, $\delta=\beta=0.5$, and $k=3, 5, 7$ for lines with
thickness increases with $k$.}
 \label{fig:VLRCaseIIk}
\end{figure}

Next we consider a simple generalization of the usual Yukawa
coupling: $F(\phi)=\phi^k$, where $k$ is a positive odd integer. We
have known in previous subsection that, for the usual Yukawa
coupling $(k=1)$, the appearance of a well in the effective
potential of right-handed fermions needs large couple constant
$\eta$, i.e., strong kink-fermion coupling. However, the case
$F(\phi)=\phi^k$ with $k>1$ has a very different characteristic from
the $k=1$ case at the brane location. This can be seen from the
derivative of $F(\phi(z))$ with respect to $z$ at $z=0$:
\begin{eqnarray}
 \partial_z F(\phi(0))=k\phi^{k-1}(0) \partial_z\phi(0)
 \bigg\{\begin{array}{c}
    \neq 0~~~\text{for}~~ k=1, \\
    =0~~~\text{for}~~ k>1.
  \end{array}
\label{dzF}
\end{eqnarray}
Together with $\phi(0)=\phi^k(0)=0$, they would result in different
values of the effective potentials of fermions (\ref{Vfermion}) at
brane location for two kind of coupling. For the current case, the
potentials (\ref{Vfermion}) are now
\begin{subequations}\label{VSLRphik}
\begin{eqnarray}
 V_L(z)
  &=&\frac{\arctan^{2k} \sinh \left( \frac{\beta z}{\delta}\right)}
           {\cosh^{2\delta} \left( \frac{\beta z}{\delta}  \right)}
      \phi_0^{2k}\eta^2 \nonumber \\
  &&    - \frac{\arctan^{k-1} \sinh \left( \frac{\beta z}{\delta}\right)}
           {\delta\cosh^{1+\delta} \left( \frac{\beta z}{\delta}  \right)}
         {k\phi_0^{k}\beta \eta} \nonumber \\
  &&  +\frac{\arctan^{k} \sinh \left( \frac{\beta z}{\delta}\right)}
           {\cosh^{\delta} \left( \frac{\beta z}{\delta}  \right)
            \coth\left( \frac{\beta z}{\delta}  \right)}
         {\phi_0^{k}\beta \eta},  \label{VSLphi} \\
  V_R(z) &=& V_L(z)|_{\eta \rightarrow -\eta}. \label{VSRphi}
\end{eqnarray}
\end{subequations}
Both potentials have same asymptotic behavior when $z \rightarrow
\pm\infty$ and $z=0$:
\begin{eqnarray}
 V_{L,R}(\pm\infty)=V_{L,R}(0)=0.
\end{eqnarray}
Therefore, for any positive $\eta$, there is a potential well for
right-handed fermion. Especially, the potential well for left-handed
fermions becomes a double-well. The shapes of the two potentials are
shown in Figs. \ref{fig:VLRCaseII} and \ref{fig:VLRCaseIIk} for
different values of $\beta$, $\delta$, $\eta$ and $k$. The depth of
the potential well increases with $\beta$, $\eta$ and $k$, and
decreases when $\delta\rightarrow 0$ or $1$. It is remarkable that,
even for a very small $\eta$, there is a potential well for both
left and right handed fermions. For a large $\eta$, the potential
bars for both fermions are almost equal. However, for a small one,
the potential bar of right-handed fermions is much higher than that
of the left-handed ones.

\begin{figure}[htb]
\begin{center}
\includegraphics[width=7cm,height=4.5cm]{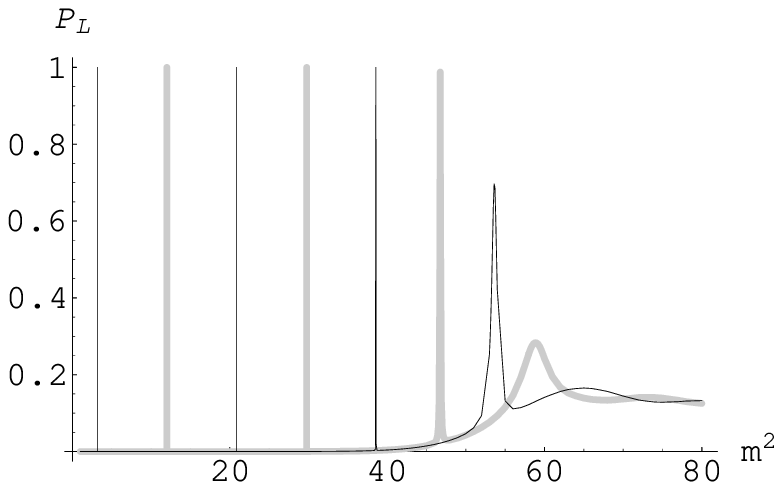}
\includegraphics[width=7cm,height=4.5cm]{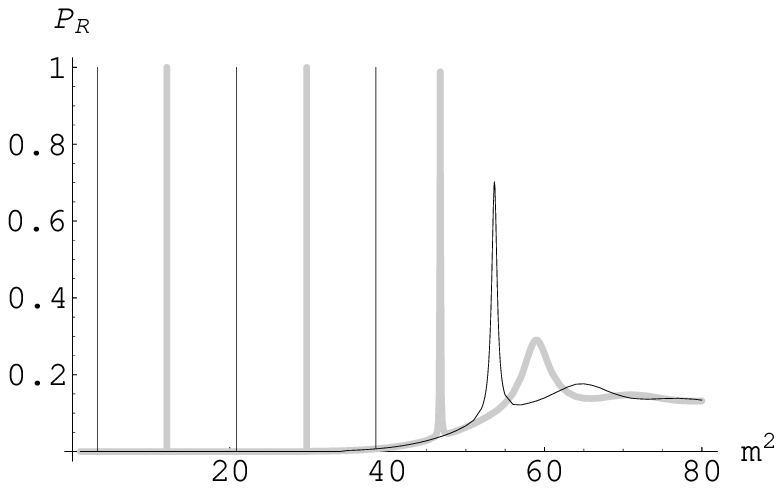}
\end{center}\vskip -4mm
\caption{The probability for finding massive KK modes of left and
right chiral fermions with mass $m^2$ around the brane location,
$P_{L,R}$, as a function of $m^2$, for the case $F(\phi)=\phi^3$.
The parameters are $\delta=\beta=0.5$ and $\eta=10$. For $P_L$,
gray thick lines and black thin lines are plotted for the first
and second initial condition, respectively. For $P_R$, black thin
lines and gray thick lines are plotted for the first and second
initial condition, respectively.}
 \label{fig:PLRphi3}
\end{figure}

There exists a continuous gapless spectrum of the KK modes for both
the left chiral and right chiral fermions. The zero mode of the left
chiral fermions
\begin{equation}
 f_{L,0}(z)  \propto \exp\left(-\eta\int^z d\bar{z}
   \text{e}^{A(\bar{z})}\phi^k(\bar{z})\right)
  \label{zeroModeL}
\end{equation}
is also non-normalizable. We also expand the effective potential
(\ref{VSRphi}) of right chiral fermions in the neighborhood of the
brane for the purpose of obtaining the resonance solutions of
massive KK modes:
\begin{eqnarray}
  V_R&=&\frac{k\beta\eta\phi_0^k}{\delta}
          \left(\frac{\beta z}{\delta}\right)^{k-1}
     -\frac{(k+2)(k+3\delta)\beta\eta\phi_0^k}{6\delta}
          \left(\frac{\beta z}{\delta}\right)^{k+1} \nonumber \\
    && + \mathcal{O}(z)^{k+3}. ~~ (k\ge 3) \label{VRLowOrder}
\end{eqnarray}
Note that another term $\eta^2\phi_0^{2k}\left(\frac{\beta
z}{\delta}\right)^{2k}$ should be included for the case $k=1$. Now
we see that the case $k=3$ is special, for which the Schr\"{o}dinger
equation with lowest order potential can be solved analytically, and
the solution is
\begin{eqnarray}
 f_{R,n} &\propto& \mathrm{e}^{-\frac{1}{2} \sqrt{a}\; z^2}
       H_n \big(a^{1/4} z\big),\\
 m_{R,n}^2&=&(1+2n)\sqrt{a},\label{mRn2phi3}
\end{eqnarray}
where $a= 3 \eta\left({
\sqrt{3(1-\delta)\delta}\;\beta}/{\delta}\right)^3$,
$n=0,1,2,\cdots$. For lowest state, we have
\begin{eqnarray}
 m_{R,0}^2&=&\sqrt{3 \eta\left({
\sqrt{3(1-\delta)\delta}\;\beta}/{\delta}\right)^3}.
\end{eqnarray}
This is very different from the case of $k=1$, where the mass of
the first resonance is decided by $m_{R0}^2 =
2\sqrt{3(1-\delta)\delta}\; \eta{\beta}/{\delta}$. Note that, we
can not get the similar approximate spectrum solution for
left-handed fermions for the case $k\geq 3$ because of the
negative derivative of $V_L(z)$ at $z=0$. But we can also expect
that the spectrum of left-handed fermions is the same as that of
the right-handed ones. This can be checked by numerical results
(see Fig. \ref{fig:PLRphi3}). The numerical calculation shows that
the spectrum (\ref{mRn2phi3}) for large coupling $\eta$ is a good
approximation. For example, for the parameters $\delta=\beta=0.5$,
$k=3$, and $\eta=10$, the resonance spectrum calculated from the
formula (\ref{mRn2phi3}) is $m_n^2=(4.4, 13.2, 22.1, 30.9, 39.7,
48.6, 57.4)$, while the numerical result is $m_n^2=(4.1, 12.3,
20.8, 29.8, 38.6, 46.8, 53.6)$. The life-times for the
$n=0,1,2,4,6$ resonances are estimated as $\tau=1.39\times
10^{10},5.81\times 10^{7},9.57\times 10^{5},1.49\times
10^{3},15.5$ respectively. It shows that the resonances with lower
$n$ would have longer life-time and may be found in future high
energy experiments. For small $\beta$ and $\eta$, the potential
has also similar well shape, but the numerical calculation shows
that there is no resonances with long life-time. This is because
the depth of the well is not deep enough.

\begin{figure}[htb]
\begin{center}
\subfigure[$k=1,n=N-1=2$] {\label{fig:fRk1}
\includegraphics[width=7cm,height=4.5cm]{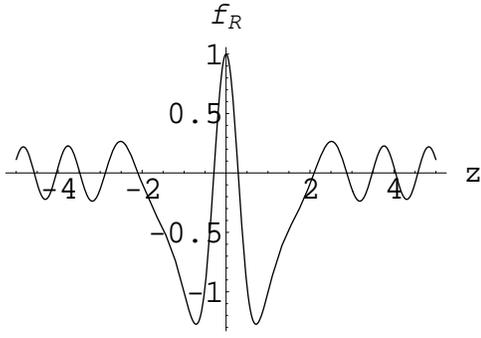}}
\subfigure[$k=3,n=N-1=6$]  {\label{fig:fRk3}
\includegraphics[width=7cm,height=4.5cm]{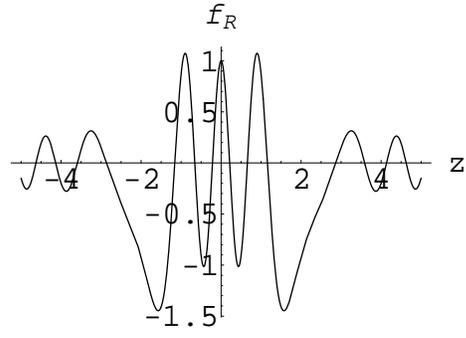}}
\subfigure[$k=5,n=N-1=12$]  {\label{fig:fRk5}
\includegraphics[width=7cm,height=4.5cm]{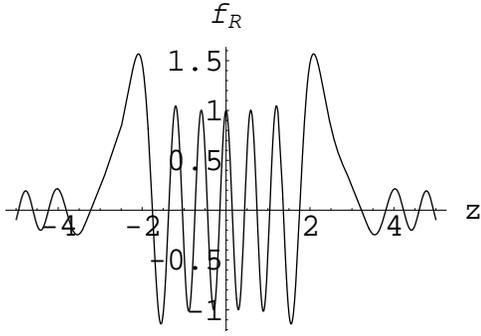}}
\subfigure[$k=7,n=N-1=23$]  {\label{fig:fRk7}
\includegraphics[width=7cm,height=4.5cm]{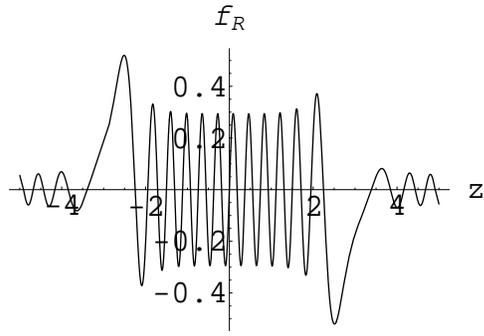}}
\subfigure[$k=9,n=N-1=43$]  {\label{fig:fRk9}
\includegraphics[width=7cm,height=4.5cm]{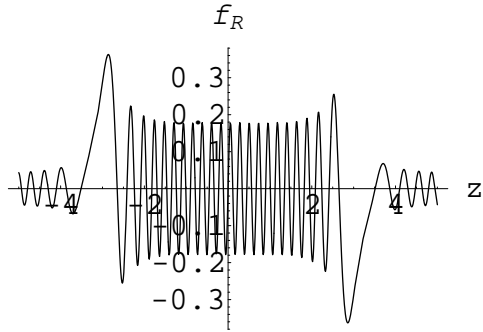}}
\end{center} \vskip -6mm
\caption{The $n=(N-1)$-th resonance modes of right chiral fermions
for the case $F(\phi)=\phi^k$. The parameters are
$\delta=\beta=0.5$, and $\eta=10$.}
 \label{fig:fRk}
\end{figure}

Comparing Fig. \ref{fig:PLR} with Fig. \ref{fig:PLRphi3}, we see
that the number of resonances of the case $F(\phi)=\phi^3$ is more
than that of the case $F(\phi)=\phi$ for a same set of parameters.
In Table \ref{tab2}, we list the relation of the number of the
resonances $N$, the mass square $m^2_{N-1}$, the probability
$P_{N-1}$, the width $\Delta m_{N-1}$ and the life-time $\tau_{N-1}$
of the $(N-1)$-th resonance of right-handed fermions with $k$ for
the set of parameters: $\delta=\beta=0.5$ and $\eta=10$. In Fig.
\ref{fig:fRk} the shapes of the $n=(N-1)$-th resonance modes of
right chiral fermions for the case $F(\phi)=\phi^k$ are plotted. The
numerical results show that the life-times of the $(N-1)$-th
resonances for different $k$ are the same order. However, the
life-time of the lower level (i.e., lower $n$) resonance for larger
$k$ is much longer than the one for smaller $k$.

\begin{table}[h]
\renewcommand\arraystretch{1.3}
\begin{center}
\caption{The relation of the number of the resonances $N$, the mass
square $m^2_{N-1}$, the probability $P_{N-1}$, the width $\Delta
m_{N-1}$ and the life-time $\tau_{N-1}$  of the $(N-1)$-th resonance
of right-handed fermions with $k$. The parameters are
$\delta=\beta=0.5$ and $\eta=10$.}\label{tab2}
\begin{tabular}
 {|c||c|c|c|c|c|}
  \hline
  $~k~$ & $~~N~~$ & $~~P_{N-1}~~$ & $~~~m^2_{N-1}~~~$
    & $~~\Delta m_{N-1}~~$ & $~~\tau_{N-1}~~$ \\
  \hline\hline
  $1$ & 3 & 0.66 & 38.84 & 0.0782 & 12.8 \\
  \hline
  $3$ & 7 & 0.70 & 53.64 & 0.0644 & 15.5 \\
  \hline
  $5$ & 13 & 0.83 & 114.1 & 0.0361 & 27.7 \\
  \hline
  $7$ & 24 & 0.37 & 287.8 & 0.0383 & 26.1 \\
  \hline
  $9$ & 44 & 0.31 & 784.0 & 0.0443 & 22.6 \\
  \hline
  $11$ & 79 & 0.33 & 2200 & 0.0171 & 58.4 \\
  \hline
\end{tabular}
\end{center}
\end{table}

\section{Discussion and conclusion}
\label{conclusion}

In this paper, by presenting the shapes of the mass-independent
potentials of KK modes in the corresponding Schr\"{o}dinger
equations, we have investigated the localization and resonance
spectrum of fermionic fields on a one-field generated $dS$ thick
brane. It is shown that, in order to localize left or right chiral
fermions on the brane, some kind of kink-fermion coupling should be
introduced. A class of kink-fermion couplings
$\eta\bar{\Psi}\phi^k\Psi$ with positive $\eta$ and positive odd
integer $k$ are investigated in detail. It is worth to point out
that, in this $dS$ thick brane scenario, the potential of KK modes
of left chiral fermions is very different from the potentials of KK
modes of the fields with spin 0, 1 and 2. For other spin fields, the
potentials of KK modes are the modified P\"{o}schl-Teller
potentials, which suggest that there exist a mass gap and a series
of continuous spectrum starting at positive $\mu^2$
\cite{WangPRD2002,LiuJCAP2009}. While the potential of KK modes of
left chiral fermions is a modified volcano type potential. The
potentials for both left and right chiral fermions vanish
asymptotically when far away from the brane, hence all values of
$\mu^2>0$ are allowed, and there exists no mass gap but a continuous
gapless spectrum of KK states with $\mu^2>0$. The massive KK modes
asymptotically turn into continuous plane waves when far away from
the brane.

For the simplest kink-fermion coupling $\eta\bar{\Psi}\phi\Psi$,
the potential of KK modes of right chiral fermions is always
positive at the brane location and vanishes when far away from the
brane. This shows that the potential could not trap any bound
fermions with right chirality and there is no zero mode of right
chiral fermions. However, with the increase of $\eta$, a potential
well around the brane location appears and the well becomes more
and more deeper. A set of massive fermions with a finite life-time
(resonances) are obtained. We find that the masses and life-times
of left and right chiral resonances are almost the same, which
demonstrates that it is possible to compose a massive Dirac
fermion from the left and right chiral resonances. This conclusion
agrees with the case of bound massive fermionic KK modes given in
Refs. \cite{Liu0803,LiuJCAP2009}, where mass spectrums of left and
right chiral massive fermions are the same.

For the case $F(\phi)=\phi^k$ with odd $k>1$, the appearance of a
well in the effective potential of right-handed fermions does not
need large couple constant $\eta$. Especially, the potential well
for left-handed fermions becomes a double-well. It is also shown
that the spectrum of left-handed fermionic resonances is the same
as that of the right-handed ones. The resonance with lower mass
has longer life-time. For a same set of parameters, the number of
resonances increases with $k$ and the life-time of the lower level
resonance for larger $k$ is much longer than the one for smaller
$k$. For small $\beta$ and $\eta$, there is no resonances with
long life-time.

At last, we give a comment on the fact that either the left or the
right resonance peaks at the brane's location itself. From the
coupled equations (\ref{CoupleEq1}), it can be seen that the massive
KK modes with different chirality but same mass are in fact
contacted with each other. Since $e^A(z) F(\phi(z))$ is odd, an odd
left-chiral massive KK mode will correspond to an even right one
with same mass, and vice versa. This results in the fact that either
the left or the right resonance peaks at the brane's location
itself. So, at the brane location $z=0$, we can not get a massive
resonance made of left and right chiral KK modes with same mass.
This is to say, if we allow measurements only at $z=0$, the massive
resonances found on the brane are all chiral, and the evidence of
Dirac fermions is lost. However, in realistic thick brane models,
branes are extended objects along the extra dimension. So we can
interpret the probability for finding the massive modes on the brane
(not necessarily at $z=0$) as $\int_{-z_b}^{z_b}dz|f_{L,R}(z)|^2$ or
more suitable as (25), where the parameter $z_b$ is chosen in order
to allow the influence of the odd modes in small regions around the
brane location $z=0$. According to this idea, the formation of
metastable massive Dirac fermions is realized, and those fermions
are quasi-localized on the brane. Similar discussions can also be
found in Refs.
\cite{Grossman2000,0901.3543,0110299,Mouslopoulos,Llatas} for
massive fermions or gravitons.

\section*{Acknowledgement}

Y.X. Liu would like to thank Adalto R. Gomes for useful discussions,
and the authors thank the referees for remarks that improved the
work. This work was supported by the Program for New Century
Excellent Talents in University, the National Natural Science
Foundation of China(NSFC)(No. 10705013), the Doctoral Program
Foundation of Institutions of Higher Education of China (No.
20070730055), the Key Project of Chinese Ministry of Education (No.
109153) and the Fundamental Research Fund for Physics and
Mathematics of Lanzhou University (No. Lzu07002).

\end{document}